%% file: sample-acmsmall.tex
\documentclass[sigconf]{acmart}

\usepackage{graphicx} 
\usepackage{color}
\usepackage[english]{babel}
\usepackage[utf8]{inputenc}
\usepackage{algorithm}
\usepackage[noend]{algpseudocode}
\usepackage{multirow}
\usepackage{multicol}
\usepackage{enumitem}
\setlist[itemize]{leftmargin=*}
\usepackage{float}
\usepackage{array}

\AtBeginDocument{%
  \providecommand\BibTeX{{%
    \normalfont B\kern-0.5em{\scshape i\kern-0.25em b}\kern-0.8em\TeX}}}

\copyrightyear{2024}
\acmYear{2024}
\setcopyright{acmlicensed}\acmConference[CIKM '24]{Proceedings of the 33rd ACM International Conference on Information and Knowledge Management}{October 21--25, 2024}{Boise, ID, USA}
\acmBooktitle{Proceedings of the 33rd ACM International Conference on Information and Knowledge Management (CIKM '24), October 21--25, 2024, Boise, ID, USA}
\acmDOI{10.1145/3627673.3679622}
\acmISBN{979-8-4007-0436-9/24/10}

\begin{document}

\title{HC-GST: Heterophily-aware Distribution Consistency based Graph Self-training}

\author{Fali Wang}
\affiliation{%
  \institution{The Pennsylvania State University}
  \city{University Park}
  \country{USA}}
\email{fqw5095@psu.edu}

\author{Tianxiang Zhao}
\affiliation{%
  \institution{The Pennsylvania State University}
  \city{University Park}
  \country{USA}}
\email{tkz5084@psu.edu}

\author{Junjie Xu}
\affiliation{%
  \institution{The Pennsylvania State University}
  \city{University Park}
  \country{USA}}
\email{jmx5097@psu.edu}

\author{Suhang Wang}
\affiliation{%
  \institution{The Pennsylvania State University}
  \city{University Park}
  \country{USA}}
\email{szw494@psu.edu}

\begin{abstract}

Graph self-training (GST), which selects and assigns pseudo-labels to unlabeled nodes, is popular for tackling label sparsity in graphs. However, recent study on homophily graphs show that GST methods could introduce and amplify distribution shift between training and test nodes as they tend to assign pseudo-labels to nodes they are good at. As GNNs typically perform better on homophilic nodes, there could be potential shifts towards homophilic pseudo-nodes, which is underexplored. 
Our preliminary experiments on heterophilic graphs verify that these methods can cause shifts in homophily ratio distributions, leading to \textit{training bias} that improves performance on homophilic nodes while degrading it on heterophilic ones. Therefore, we study a novel problem of reducing homophily ratio distribution shifts during self-training on heterophilic graphs. A key challenge is the accurate calculation of homophily ratios and their distributions without extensive labeled data. To tackle them, 
we propose a novel Heterophily-aware Distribution Consistency-based Graph Self-Training (HC-GST) framework, which estimates homophily ratios using soft labels and optimizes a selection vector to align pseudo-nodes with the global homophily ratio distribution. Extensive experiments on both homophilic and heterophilic graphs show that HC-GST effectively reduces training bias and enhances self-training performance.

\end{abstract}



\received{20 February 2007}
\received[revised]{12 March 2009}
\received[accepted]{5 June 2009}

\maketitle

\section{Introduction} 

Graphs are pervasive in the real world, such as in social networks \cite{peng2022reinforced}, knowledge graphs \cite{vashishth2019composition}, and traffic networks \cite{wu2020connecting}. Graph Neural Networks (GNNs) have achieved significant advances in semi-supervised node classification across both homophilic and heterophilic graphs \cite{kipf2017semisupervised,  abu2019mixhop, MAURYA2022101695, he2022block, chien2020adaptive, xu2023node}. Despite their success, they often rely on abundant labeled data \cite{ding2020graph, wang2023few} for training, yet manual labeling is not only labor-intensive but also impractical for new classes with sparse samples \cite{wang2022faith}. This has spurred increasing interest in node classification with sparse labels \cite{sun2020multi, liu2022confidence}.

Graph self-training (GST) emerges as a promising method to harness abundant unlabeled nodes alongside a limited number of labeled nodes to tackle label sparsity in graphs \cite{li2018deeper,lee2013pseudo}. It typically involves three core steps: (\textbf{i}) \textit{pseudo-node selection}, selecting high-confidence nodes; (\textbf{ii}) \textit{pseudo-label assignment}, assigning the most probable labels; and (\textbf{iii}) \textit{retraining}, where the GNN is trained on the augmented set, iterating these steps until convergence or a specified number of stages. However, typical confidence-based pseudo-node selection can cause distribution shifts \cite{zhu2021shift, liu2022confidence}, potentially undermining self-training. This occurs as an increasing number of easy nodes, identified by their high confidence, are added to the original labeled training set, leading to the distribution gradually shifting towards this augmented training set and excessive focus on such easy nodes as a result.
Methods like DRGST \cite{liu2022confidence} and DCGST \cite{wang2024DCGST} address this by quantifying distribution shifts through information gain and GNN representation, respectively, and selecting pseudo-nodes that align with the distribution of the unlabeled set. 
Nevertheless, these methods are tailored for homophilic graphs, where connected nodes tend to have similar labels. In contrast, they neglect shifts in neighboring distributions (homophily ratio) in heterophilic graphs, where GNNs perform well on homophilic nodes but struggle with heterophilic ones, resulting in novel distribution shifts.


To investigate self-training on heterophilic graphs, our preliminary experiments on the representative Chameleon graph \cite{Rozemberczki2021} in  Sec.\ref{sec:pre_exp_training_bias_issue} 
show that: (1) Training sets aligned with the true homophily distributions yield the best performance, as shown in Fig.\ref{fig:intro_four_fig} (a); (2) GNNs excel on homophilic nodes, with accuracy positively correlating with the homophily ratio, evident from the blue line in Fig.\ref{fig:intro_four_fig} (c). (3) Traditional self-training (ST) tends to select pseudo-nodes with higher homophily ratios, as depicted in Fig.\ref{fig:intro_four_fig} (b), resulting in distribution shifts; (4) Strategies based on confidence or GNN representation consistency often lead to training bias, improving performance on homophilic nodes but worsening it on heterophilic nodes, as illustrated in Fig.\ref{fig:intro_four_fig} (c) and (d). These results underscore the need for maintaining consistency in homophily ratio distribution (node counts across homophily bins) to mitigate training bias and improve self-training, a gap not fully addressed by existing methods. Thus, ensuring consistency between the homophily ratio distribution of the local (pseudo-)labeled training set and the global heterophilic graph to reduce training bias across homophily bins remains an open question.


Therefore, in this paper, we study a novel problem of homophily ratio distribution consistency during self-training on heterophilic graphs. However, this presents challenges to maximizing self-training performance in obtaining accurate homophily ratios and distributions, aligning homophily ratios between (pseudo-)labeled nodes and the true distribution, and the low pseudo-labeling accuracy on heterophilic nodes. Thus, we pose a critical \textbf{research question}: {\textit{How can we develop a GST framework which ensures selected pseudo-nodes align with true homophily ratio distributions and assigns accurate pseudo-labels to highly heterophilic nodes?}}

To address these challenges, we propose a novel \textbf{H}eterophily-aware \textbf{D}istribution \textbf{C}onsistency-based \textbf{G}raph \textbf{S}elf-\textbf{T}raining (HC-GST) framework that adapts distribution consistency strategies from homophilic graphs.
(1) Given the difficulty of acquiring exact homophily ratios and distributions without labels for a large number of unlabeled nodes, we develop an estimation method utilizing soft labels, which are more informative than one-hot labels.
(2) To align pseudo-nodes with the target homophily ratio distribution, we optimize a selection vector $\mathbf{q}$, each element representing the selection probability for nodes in a high-quality candidate set, selecting the top $K$ nodes as pseudo-nodes.
(3) After selecting pseudo-nodes, our framework assigns pseudo-labels. However, nodes with high heterophily often show lower accuracy because local neighbors typically belong to different classes. Inspired by \cite{abu2019mixhop, zheng2022graph}, we employ multi-hop neighbors to enhance labeling accuracy for heterophilic nodes, while using one-hop neighbors for homophilic nodes.
Furthermore, although we carefully select pseudo-nodes that align with the target distribution, this approach may exclude inconsistent yet potentially high-quality nodes. To fully utilize these nodes, we introduce a dual-head GNN model: the main classifier head trains on clean and selected pseudo-nodes, while an auxiliary head focuses on the previously discarded high-quality pseudo-nodes, optimizing the feature extractor without compromising the performance of the main classifier.
Our main contributions are:
\begin{itemize}[itemsep=0pt,topsep=0pt,parsep=0pt,leftmargin=*]
    \item We study a novel problem of reducing training bias across homophily bins caused by shifts in homophily ratio distribution, unique in heterophilic graphs compared to homophilic graphs.
    \item We address the challenges of adapting distribution-consistent graph self-training methods to heterophilic graphs by introducing the innovative HC-GST framework, which utilizes a homophily ratio distribution consistency selection strategy.
    \item Extensive experiments on both homophilic and heterophilic graphs consistently demonstrate the effectiveness of HC-GST in reducing training bias.
\end{itemize}

\section{Related Work}



\noindent\textbf{Graph Neural Networks with Heterophily}. 
Graph Neural Networks (GNNs) update node representations through neighbor aggregation, aiding tasks like node classification~\cite{jin2021heterogeneous, hu2019hierarchical}. Their performance drops in heterophilic graphs~\cite{wu2019simplifying, chen2018fastgcn}. Solutions include higher-order neighbor aggregation (H2GCN~\cite{zhu2020beyond}, MixHop~\cite{abu2019mixhop}) and differentiated message passing (GGCN~\cite{yan2022two}, GPR-GNN~\cite{chien2020adaptive}). BMGCN~\cite{he2022block} uses a block similarity matrix to adapt to both graph types.

\noindent\textbf{Graph Self-training}. 
Training GNNs with sparse labels for node classification often leverages self-training, using vast unlabeled data to enhance performance~\cite{scudder1965probability, zoph2020rethinking, lee2013pseudo, mukherjee2020uncertainty}. Self-training expands the training set with high-quality pseudo-labels~\cite{li2018deeper}, primarily using confidence-based selection~\cite{sun2020multi, zhou2019effective, li2018deeper, wang2021confident, zoph2020rethinking}. Techniques like M3S~\cite{sun2020multi}, which integrates deep clustering, and DSGSN~\cite{zhou2019effective}, using negative sampling, address GNN training instability. However, methods like DRGST~\cite{liu2022confidence} and DCGST~\cite{wang2024DCGST} that adjust to distribution shifts still assume homophily, limiting their effectiveness in heterophilic graphs. This paper adapts distribution consistency to better suit heterophilic graph characteristics.

\vskip 0.3em

\noindent\textbf{Distribution Shifts on Graphs}. Training and testing nodes are assumed to have consistent distributions in most node classification benchmarks. However, real-world applications often experience distribution shifts between the labeled training sets and the actual data \cite{wu2022recent}, leading to GNN classifiers overfitting, and negatively impacting deployment performance. Domain-Invariant Representation (DIR) learning addresses this by minimizing distribution differences across domains; using adversarial learning \cite{ganin2016domain}, or heuristic measures such as CMD \cite{zellinger2019robust} and MMD \cite{long2017deep}. In self-training, inappropriate pseudo-node selection exacerbates distribution shifts \cite{liu2022confidence, wang2024DCGST}. The unique challenge of homophily ratio distribution shifts in heterophilic graphs remains unexplored. This study is the first to explore shifts in homophily ratio distribution in heterophilic graphs within a self-training framework.

\section{Preliminary}
In this section, we discuss self-training methods that tackle label sparsity and distribution shifts in homophilic graphs. Additionally, we conduct preliminary experiments to validate the unique training bias during self-training on heterophilic graphs. Finally, we formalize our problem definition.

\subsection{Notations and GNNs} 
We use $\mathcal{G}=\{\mathcal{V}, \mathcal{E}, \mathbf{X}\}$ to denote an attributed graph with $\mathcal{V}=\{v_1, \dots, v_n\}$ as the set of nodes, $\mathcal{E}$ as the set of edges, and $\mathbf{X}$ as a feature matrix of $d$-dimensional node features. The adjacency matrix $\mathbf{A}$ indicates connections with $\mathbf{A}_{ij}=1$ if $v_i$ and $v_j$ are connected, otherwise $0$. In semi-supervised node classification, only a small portion of nodes, $\mathcal{V}^L$, have labels $\mathbf{Y}^L$ from a one-hot matrix $\mathbf{Y}$. The goal is to predict labels for unlabeled nodes using $\mathcal{G}$ and $\mathcal{V}^L$.

Graph neural networks have shown great ability in modeling graphs. Generally, GNNs update node representations through a message-passing mechanism, aggregating features from neighboring nodes. Let $\mathbf{Z}_i^{(l)}$ denote $v_i$'s representation at the $l$-th GNN layer. Then the message passing can be expressed as
\[
\mathbf{Z}_i^{(l+1)} = \text{Aggregate}\Big(\mathbf{Z}_i^{(l)}, \sum_{v_j \in \mathcal{N}(v_i)} \text{Propagate}\big(\mathbf{Z}_i^{(l)}, \mathbf{Z}_j^{(l)}, \mathbf{A}_{i,j}\big)\Big).
\]
where $\mathcal{N}(v_i)$ denotes the set of neighbors of $v_i$. We denote $\mathbf{Z}$ as the node representations in the last GNN layer. 

\subsection{Pseudo-labeling and Distribution Shift} 

Graph self-training is a widely used framework to tackle label sparsity by effectively leveraging large amounts of unlabeled data for performance gains. The success of self-training is its pseudo-labeling strategy, which carefully expands the training set with high-quality pseudo-nodes. However, graph data often exhibit distribution shifts that challenge self-training and pseudo-labeling strategies: 
(i) Label sparsity typically results in a non-representative training set, leading to distribution shifts away from the true distribution. These shifts can undermine models based on the IID assumption, as directly minimizing the average loss during training fails to yield predictors that generalize well under test distribution changes \cite{li2022out}. This prevents selecting high-quality pseudo-nodes; and 
(ii) An unrepresentative training set also causes imbalances in model capabilities, which during pseudo-label generation, tend to favor easier nodes. Consequently, this preference leads to an expanded pseudo-node set that deviates from the true distribution, adversely impacting self-training performance. 
Given the complex structure of graphs, measuring distribution shifts is challenging. To address this, \citet{zhu2021shift} has defined distribution shifts within the GNN representation space as the Central Moment Discrepancy (CMD) \cite{zellinger2019robust} distance between the global node representations $\mathbf{Z}^G$ from GNNs and their labeled counterparts $\mathbf{Z}^L$:  
\begin{equation}
\small
\begin{aligned}
\text{CMD}(\mathbf{Z}^G, \mathbf{Z}^L) & = \frac{1}{|b-a|}  \left\| E(\mathbf{Z}^G)-E(\mathbf{Z}^L) \right \|_2 \\
& +\sum_{k=2}^\infty \frac{1}{|b-a|^k} \left\| c_k(\mathbf{Z}^G)-c_k(\mathbf{Z}^L)\right \|_2
\label{eq:cmd}
\end{aligned}
\end{equation} 
where $E$ denotes the expectation, $c_k$ is the $k$-th order moment, and $a, b$ are the joint distribution supports, typically calculating only up to $k=5$ moments.

\subsection{Training Bias of GST in Heterophilic Graphs}
\label{sec:pre_exp_training_bias_issue}
\begin{figure}[!t]
    \centering
    \includegraphics[width=0.8\linewidth]{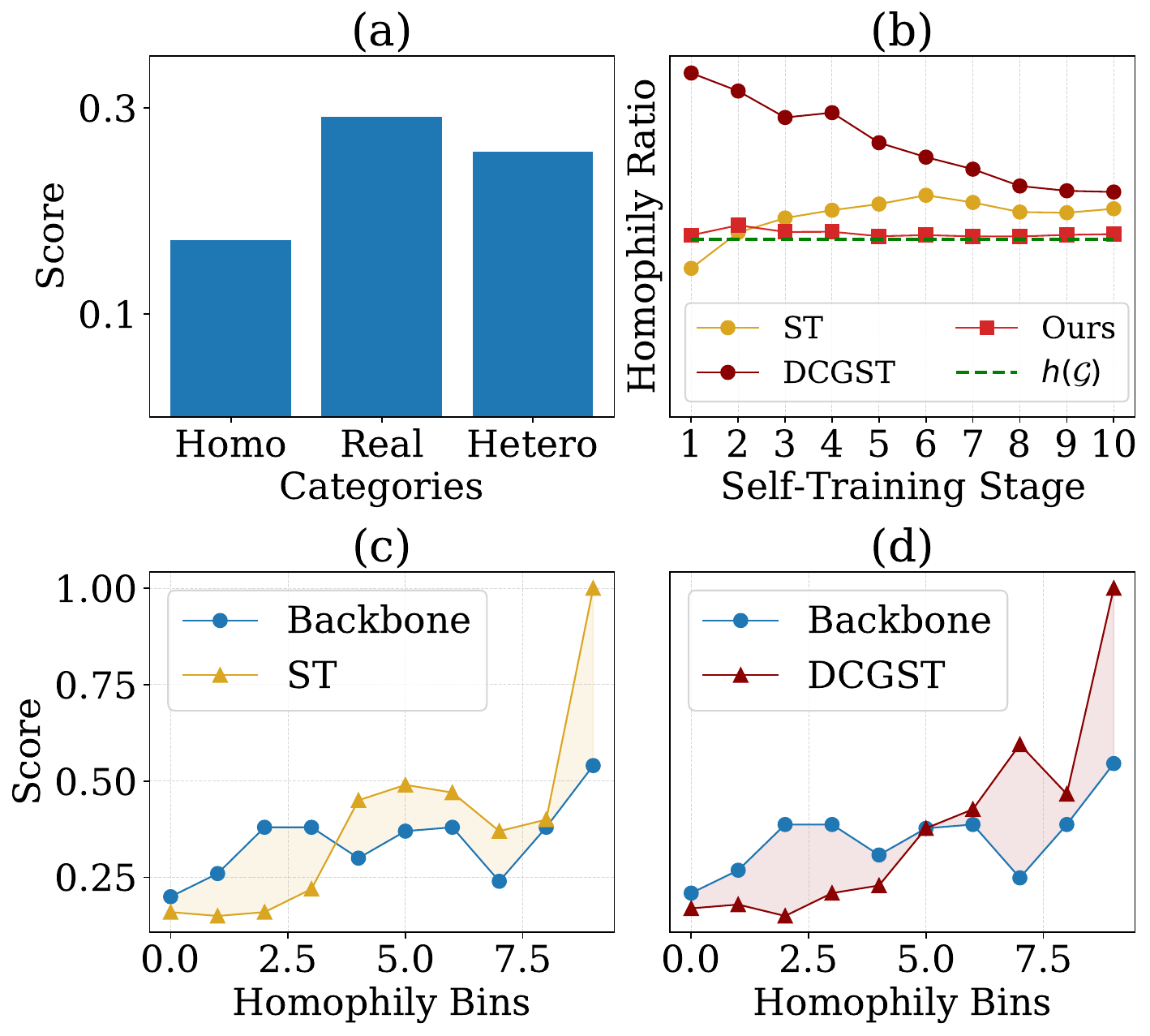}
    \vskip -1em
    \caption{Preliminary experiments of BMGCN \cite{he2022block} on the Chameleon Graph~\cite{Rozemberczki2021}. (a) Performance across crafted training sets biased towards homophily, aligned with the true graph, and biased towards heterophily. (b) Variation in mean homophily ratio across self-training stages for the BMGCN backbone, ST, and DCGST. (c) Training biases in traditional graph self-training (ST) shown through accuracies across different homophily bins. (d) Training biases in DCGST \cite{wang2024DCGST}. }
    \vskip -1.2em
    \label{fig:intro_four_fig}
\end{figure}

Existing work reduces distribution shifts in homophilic graphs within self-training frameworks \cite{liu2022confidence, wang2024DCGST}. However, these methods do not extend well to heterophilic graphs, which may face unique challenges due to shifts in heterophily-aware distributions. To explore the unique challenges of self-training on heterophilic graphs, we conduct preliminary experiments. Heterophilic graphs mainly differ from homophilic ones in that a node tends to connect to nodes of dissimilar features or classes. To measure this, we first define node and graph homophily ratios.
\vskip -1em
\begin{definition}[Node Homophily Ratio $h(v_i)$~\cite{pei2020geom}] $h(v_i)$ for a node $v_i$ measures the ratio of $v_i$'s neighbors that share the same label as $v_i$, which can be written as $h(v_i) = \frac{|\{ v_j \in \mathcal{N}_i : y_j = y_i \}|}{|\mathcal{N}_i|}$, 
where $\mathcal{N}_i$ means the neighbors of node $v_i$ and $y_i$ is its golden label. 
\end{definition}
\vskip -1em
\begin{definition}[Graph Homophily Ratio $h(\mathcal{G})$~\cite{pei2020geom}] $h(\mathcal{G})$ represents the average node homophily ratio across all nodes in the graph $\mathcal{G}$, which is calculated as $h(\mathcal{G}) = \frac{1}{|\mathcal{V}|} \sum_{v_i \in \mathcal{V}} h(v_i)$.
\end{definition}
To quantify the distribution of homophily ratios, we evenly divide the homophily level into $N$ bins as
\begin{equation}
    \mathcal{B} = \left [ |\mathcal{B}_1|, |\mathcal{B}_2|, \dots, |\mathcal{B}_N| \right ]
    \label{eq:homo_ratio_distribution}
\end{equation}
where 
\begin{equation}
    \mathcal{B}_i = \left\{ v \mid h(v) \in \left[\frac{i-1}{N}, \frac{i}{N}\right) \right\}
\end{equation}

The \textit{global homophily ratio distribution} $\mathcal{B}^G$ reflects these counts across the entire graph, whereas the \textit{local homophily ratio distribution} $\mathcal{B}^L$ refers only to the (pseudo-)labeled nodes.

Building on these concepts, we conduct preliminary experiments with the state-of-the-art heterophilic GNN model BMGCN \cite{he2022block} and the representative heterophilic graph Chameleon~\cite{Rozemberczki2021}. Initially, we evaluate the backbone GNN performance under the heterophily setting.
(1) \textbf{Impact of Homophily Ratio Distribution on Training Set}: We construct training sets based on the known global homophily ratio distribution as: biased towards homophily, representative global distributions, and towards heterophily. 
This construction selects nodes until each homophily bin $\mathcal{B}_i$ meets its target. In homophily-biased sets, higher-index bins have more nodes (e.g., $\sum_{i\in [N-3,N]}|\mathcal{B}_{i}| = |\mathcal{V}^L|$); in heterophily-biased sets, lower-index bins are denser (e.g., $\sum_{i\in [1,4]}|\mathcal{B}_i| = |\mathcal{V}^L|$). For representative sets, bin sizes align with the global distribution, defined as $|\mathcal{B}^{\text{target}}_i| = \frac{|\mathcal{B}_i^G|}{\sum_i |\mathcal{B}_i^G|} \times |\mathcal{V}^L|$.
As shown in Fig.\ref{fig:intro_four_fig} (a), sets aligned with the true distribution perform best. Compared to homophily-biased sets, those biased towards heterophily more closely mirror the true distribution, thereby achieving better performance.
(2) \textbf{Performance Analysis Across Homophily Bins}: We display the accuracy for each homophily bin (illustrated by the blue line in Fig. \ref{fig:intro_four_fig} (c)). We find that accuracy positively correlates with the homophily ratio, confirming that GNNs are inherently better at processing homophilic nodes than heterophilic ones.

Next, we continue to explore the effects of heterophily within the self-training framework.
(3) \textbf{Change of Homophily Ratio under Self-Training}: We analyze homophily ratio changes under two self-training frameworks: traditional confidence-based graph self-training (ST) \cite{li2018deeper} and distribution-consistent graph self-training (DCGST) \cite{wang2024DCGST}. We calculate average homophily ratios for (pseudo-)labeled nodes (shown in Fig.\ref{fig:intro_four_fig} (b)). Results show ST selects nodes with higher homophily, while DCGST reduces the average but still deviates from the global average, indicating previous self-training methods struggle to align well with global homophily ratios. We refer to this as the Homophily Ratio Distribution Shift. 
(4) \textbf{Training Bias induced by Homophily Ratio Distribution Shift}: To assess the impact of shifts in homophily ratio distribution, we analyze the accuracy of both self-training methods across various homophily bins, as shown in Fig.\ref{fig:intro_four_fig} (c) and (d). Our findings indicate that the model excels in homophily-biased bins, where it naturally performs well, and struggles with heterophilic nodes, akin to the Matthew Effect in economics. We describe this phenomenon as training bias across homophily bins.
After identifying unique distribution shift issue across homophily bins in heterophilic graphs, we define it as:
\vskip -1em
\begin{definition}[Distribution Shift on Heterophily]
Assume the \textit{global homophily ratio distribution} $\mathcal{B}^G$ represents the count of nodes within each homophily bin across the entire graph, and the \textit{local homophily ratio distribution} $\mathcal{B}^L$ applies to the locally labeled node set. Distribution shift $D(\mathcal{B}^G, \mathcal{B}^L)$ is measured using metrics such as KL divergence or CMD between $\mathcal{B}^G$ and $\mathcal{B}^L$.
\label{def:distribution_shift_heterophily}
\end{definition}

\subsection{Problem Definition}

With the above analysis, our problem is formally defined as: Given a graph $\mathcal{G}$ with a small set of labeled nodes $\mathcal{V}^L$ and a large number of unlabeled nodes $\mathcal{V}^U$, self-training may induce or amplify the distribution shifts in homophily ratios between the labeled node set $\mathcal{V}^L \cup \mathcal{V}^P$ and the global graph. 
Our objective is to develop a graph self-training framework that enables a GNN model $f_\theta$ to accurately predict $\hat{\mathbf{y}}_i = \arg\max_j f_\theta(v_i)_j$ for each unlabeled node, while reducing training biases through the following goals:
(1) Maximize positive performance variation: $\max_{\theta} \sum_{\Delta_i > 0} \Delta_i,
$
where $\Delta_i$ represents the performance variance in the $i$-th homophily bin $\mathcal{B}_i$.
(2) Minimize negative performance variation: $\min_{\theta} \sum_{\Delta_i < 0} |\Delta_i|$.
(3) Overall, maximize performance variation: $\max_{\theta} \sum_{i}^{N} \Delta_i$.

\input{methodology}

\input{experiment}

\section{Conclusion}


In this study, we address a unique self-training problem on heterophilic graphs: training bias due to shifts in homophily ratio distribution. We introduce HC-GST, a self-training framework designed for homophily ratio distribution consistency, which selects pseudo-nodes that align with the global using soft labels to estimate homophily ratios. Utilizing multi-hop neighbors for accurate pseudo-labeling and a dual-head GNN to fully exploit pseudo-labels further enhances model performance. To measure training bias on heterophilic graphs, we introduce three new metrics: $TPV$, $NPV$, and $PPV$. Extensive experiments on both homophilic and heterophilic graphs consistently confirm HC-GST's effectiveness in reducing training bias and improving self-training performance.

\section*{ACKNOWLEDGEMENTS}
This material is based upon work supported by, or in part by, the National Science Foundation (NSF) under grant number IIS-1909702, Army Research Office (ARO) under grant number W911NF-21-10198, Department of Homeland Security (DHS) CINA under grant number 17STCIN00001-05-00, and Cisco Faculty Research Award.


\bibliographystyle{ACM-Reference-Format}
\bibliography{sample-base}


\end{document}

%% file: methodology.tex
\section{Proposed Method}

\begin{algorithm}[t]
\caption{HC-GST Framework Workflow}
\vskip -0em 
\label{algo}
\begin{algorithmic}[1]
\Require $\mathbf{A}, \mathbf{X}, \mathcal{V}^L$, confidence threshold $\delta_c$, number of new pseudo-nodes $K$ to select in each stage, maximum number of stages $S$
\Ensure Learned classifier $f_\theta$ and its output $\mathbf{Z}$

\State $s \gets 1$  \Comment{Initialize stage counter}
\State $f_\theta \gets \text{Train}(\mathcal{V}^L, \mathbf{A}, \mathbf{X})$ \Comment{Train an initial GNN on $\mathcal{V}^L$}
\State $\mathbf{Z} \gets f_\theta(\mathbf{A}, \mathbf{X})$ \Comment{Get node representations}
\State $\mathcal{V}_{*,\leq s-1}^P \gets \emptyset$  \Comment{Initialize pseudo-label set as empty}

\Repeat
    \Statex  $//${Pseudo-node Selection} 
    \State $\mathcal{C} \gets \{ v_i \mid \sigma(\mathbf{z}_i) > \delta_c \text{ and } v_i \notin \mathcal{V}_{*,\leq s-1}^P \}$ \Comment{Form $\mathcal{C}$}
    \State $\mathbf{q} \gets \text{Optimize}(\mathcal{C}, \mathbf{Z}, \mathcal{B})$ \Comment{Optimize selection vector}
    \State $\mathcal{V}_{*,s}^P \gets \text{TopK}(\mathbf{q}, K, \mathcal{C})$ \Comment{Select top-K pseudo-nodes}
    \Statex $//${Pseudo-labeling Using Multi-hop Neighbors}
    \State $\mathbf{Z}' \gets \text{Mixing}(\mathcal{V}_{*,s}^P, \mathbf{Z}, \mathbf{H})$ \Comment{Apply multi-hop neighbors}
    \State $\mathbf{y}_i^{\text{pl}} \gets \arg \max_j \mathbf{Z}_{ij}'$ \Comment{Assign pseudo-labels}
    \Statex $//${Utilization of Pseudo-nodes via Dual-head GNN}
    \State $\mathcal{V}_{*, \leq s}^P \gets \mathcal{V}_{*, \leq s-1}^P \cup \mathcal{V}_{*,s}^P$ \Comment{Update pseudo-labeled set}
    \State $f_\theta \gets \text{Train}(\mathcal{V}^L \cup \mathcal{V}_{*, \leq s}^P, \mathbf{A}, \mathbf{X})$ \Comment{Retrain dual-head GNN}
    \State $\mathbf{Z} \gets f_\theta(\mathbf{A}, \mathbf{X})$ \Comment{Update node representations}
    \State $s \gets s + 1$  \Comment{Increment stage counter}
\Until{$s > S$ or convergence}
\State \Return Best $f_\theta$, $\mathbf{Z} \gets f_\theta(\mathbf{A}, \mathbf{X})$
\end{algorithmic}
\end{algorithm}

\begin{figure*}
    \centering
    \includegraphics[width=0.9\textwidth]{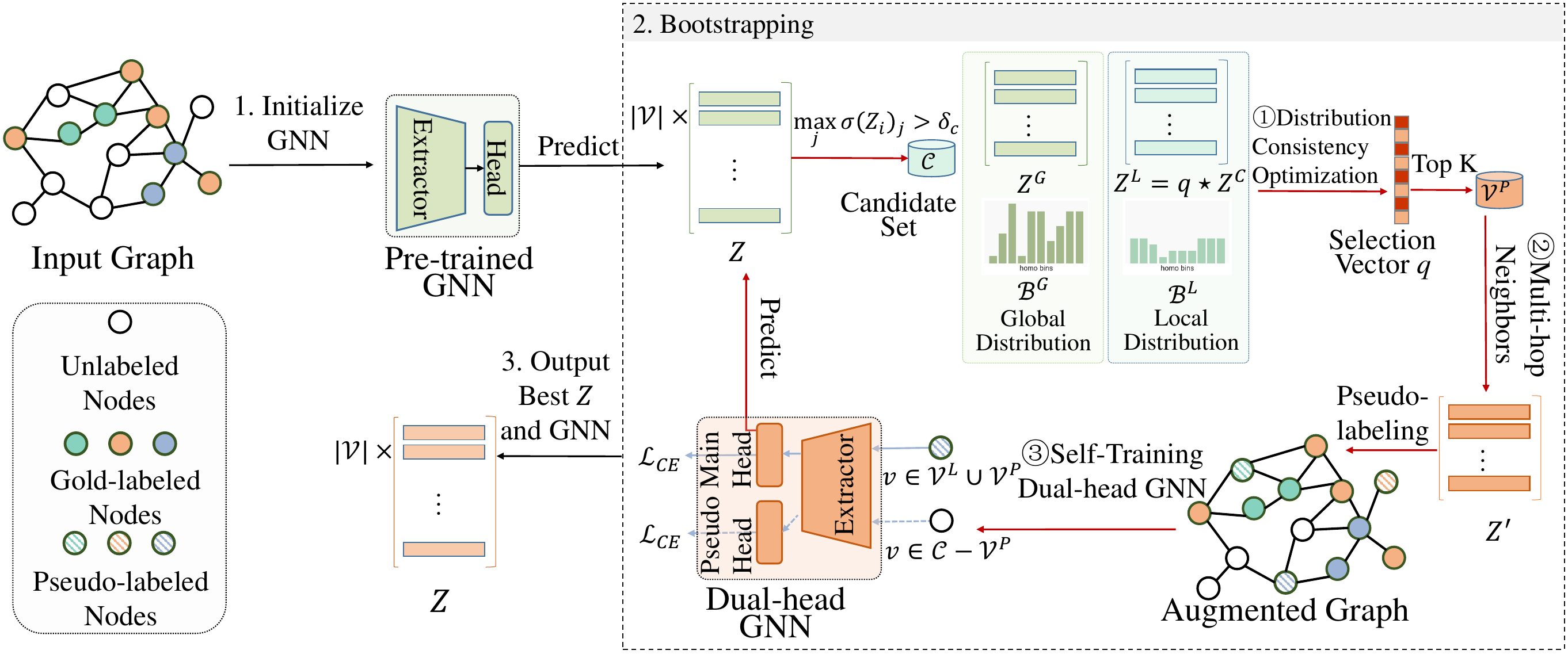}
    \vskip -1em
    \caption{The heterophily-aware distribution consistency-based graph self-training framework. Red arrows indicate the loop.}
    \vskip -1em 
    \label{fig:framework}
\end{figure*}

In this section, we give the details of the proposed framework \textbf{H}eterophily-aware Distribution \textbf{C}onsistency based \textbf{G}raph \textbf{S}elf-\textbf{T}raining (\textbf{HC-GST}). The core design of our proposed HC-GST is shown in Fig. \ref{fig:framework} and Algo. \ref{algo}, which targets heterophilic structures in three main steps: selecting high-quality pseudo-nodes that align with the true homophily ratio distribution, pseudo-labeling them using multi-hop neighbors, and utilizing them by a dual-head GNN. Specifically, the self-training framework begins by carefully selecting pseudo-nodes aligned with the global homophily ratio distribution to avoid training bias. 
Then, the step of assigning pseudo-labels improves the labeling accuracy of selected heterophilic nodes by utilizing multi-hop neighbors, reducing noise in the self-training.
Finally, we differentiate training strategies between carefully selected pseudo-nodes and the rest, to optimize the use of all potentially high-quality pseudo-nodes. Next, we detail the HC-GST framework.


\subsection{Heterophily-aware Distribution Consistent Pseudo-node Selection}
GNNs typically perform better on homophilic nodes, as shown in Fig. \ref{fig:intro_four_fig} (c). This leads to a shift in the \textit{homophily ratio distribution} during confidence-based pseudo-node selection, as self-training tends to favor homophilic nodes, thus deviating from the actual data distribution. Particularly in heterophilic graphs, this shift enhances performance on easier homophilic nodes but degrades the performance on harder heterophilic nodes, introducing a unique \textit{training bias} across homophily bins, as highlighted in Fig. \ref{fig:intro_four_fig} (c) and (d). 
To counter this training bias, we design a Heterophily-aware Distribution Consistent Pseudo-node Selection strategy, as detailed in lines 6 to 8 of Algo. \ref{algo}.  
This strategy includes three core components: (i) estimating homophily ratios and distributions to align local and global distributions; (ii) identifying target distributions to ensure new pseudo-nodes help compensate for the \textit{homophily ratio distribution shift} between local and global; and (iii) optimizing a selection vector to meet distribution consistency criteria.

\subsubsection{Homophily Ratios and Distributions Estimation}
To ensure distribution consistency in homophily ratios with the global graph during pseudo-node selection, we precisely define the \textit{Homophily Ratio Distribution} in Eq. \ref{eq:homo_ratio_distribution}, using histogram techniques \cite{bray2022evaluation}. However, calculating homophily ratios is challenging due to unknown node labels. Inspired by knowledge distillation techniques \cite{hinton2015distilling}, which suggest soft labels are more informative than one-hot labels, we estimate homophily ratios using soft label similarity as:
\begin{align}
\hat{h}(v_i) = \frac{\sum_{v_j \in \mathcal{N}_i} S(\tilde{\mathbf{y}}_j, \tilde{\mathbf{y}}_i)}{|\mathcal{N}_i|}, \quad
S(\tilde{\mathbf{y}}_j, \tilde{\mathbf{y}}_i) = \frac{\tilde{\mathbf{y}}_j \cdot \tilde{\mathbf{y}}_i}{\|\tilde{\mathbf{y}}_j\| \|\tilde{\mathbf{y}}_i\|},
\end{align}
where $\tilde{\mathbf{y}}_i$ is the soft label vector for node $v_i$ from the backbone GNN. This estimation updates as the model improves during self-training. This approach enables us to identify nodes within specific homophily ratio bins $\mathcal{B}_i$ and adjust the local distribution $\mathcal{B}^L$.

\subsubsection{Target Homophily Ratio Distribution.} 
With the estimated global distribution $\mathcal{B}^G$ and local distribution $\mathcal{B}^L$, our objective is to minimize the distance between these distributions by strategically selecting local pseudo-nodes to follow $\mathcal{B}^G$. This minimization involves adding a predefined number, $K$, of new pseudo-nodes in each stage to the local distribution to better align $\mathcal{B}^L$ with $\mathcal{B}^G$:
\[
\min_{\text{pseudo-nodes}} d(\mathcal{B}^L, \mathcal{B}^G)
\]
where $\mathcal{B}^L$ is updated by including $K$ new pseudo-nodes that optimally match $\mathcal{B}^G$. This approach ensures that the expanded set of labeled nodes in $\mathcal{V}^L \cup \mathcal{V}_{*, \leq s-1}^P$ (where $\mathcal{V}_{*, \leq s-1}^P$ includes previously added old pseudo-nodes before the current state $s$, and $\mathcal{V}_{*, s}^P$ represents the new pseudo-nodes to be added in the current stage $s$) yields a local distribution that is representative of the global distribution, effectively reducing distribution shifts caused by initial node and subsequent pseudo-node selection. Specifically, we denote $fr_i$ as the frequency rate of nodes within the $i$-th bin of global distribution $\mathcal{B}_i^G$, and $(\mathcal{V}^L \cup \mathcal{V}_{*, \leq s-1}^P)_i$ as the count of nodes in the $i$-th local bin prior to selecting new pseudo-nodes. 
The target number of nodes in the $i$-th bin is computed as follows:
\begin{align}
\small
    |\mathcal{B}_i^{\text{target}}| &= \max \left (\left\lceil fr_i \cdot (K + |\mathcal{V}^L \cup \mathcal{V}_{*, \leq s-1}^P|) - |(\mathcal{V}^L \cup \mathcal{V}_{*, \leq s-1}^P)_i| \right\rceil, 0 \right ) \notag \\
    \mathcal{B}^{\text{target}} &= \left [ |\mathcal{B}_1^{\text{target}}|, |\mathcal{B}_2^{\text{target}}|, \dots, |\mathcal{B}_N^{\text{target}}| \right ]
    \label{eq:target_distribution}
\end{align}
The fundamental principle is that local homophily ratio distributions may not align with the global, so we calculate the difference between global and local homophily ratios to set node numbers per bin. This compensates the local homophily distribution to match the global, setting the target number of pseudo-nodes per bin.


\subsubsection{Pseudo-node Selection in Heterophilic Graphs.} 
Next, we identify pseudo-nodes that align with the target homophily ratio distribution $\mathcal{B}^{\text{target}}$. However, this target distribution does not account for the distribution shifts in node representations that is helpful for maintaining consistency in graph structures. To address this, we employ the CMD metric, a popular method for measuring distribution distances in GNN representations. This metric helps us ensure consistency between the local node representations $\mathbf{Z}^L$ and the global representations $\mathbf{Z}^G$, where $\mathbf{Z}$ represents the node outputs from the backbone GNN.
Thus, we build two selection criteria: (1) selecting nodes that conform to the target homophily ratio distribution and (2) minimizing the CMD distance between the local $\mathbf{Z}^L$ and global $\mathbf{Z}^G$.
To implement this, we initialize a selection vector $\mathbf{q}$, where each element $\mathbf{q}_i \in [0, 1]$ represents the selection probability of the $i$-th node in the high-quality candidate pseudo-node set $\mathcal{C}$. 
The candidate pseudo-node set $\mathcal{C}$ is defined as follows:
\begin{equation}
\mathcal{C} = \left \{ v_i \mid \max_j \sigma(\mathbf{Z}_i)_j > \delta_c \text{ and } v_i \notin \mathcal{V}^P_{*,\leq s-1} \right \}
\label{eq:definition_candidate}
\end{equation}
where $\sigma$ is the softmax function, and $\delta_c$ represents the confidence threshold. This set includes nodes whose maximum softmax scores exceed $\delta_c$, excluding those already selected in previous stages. This selective criterion ensures that the candidate set $\mathcal{C}$ contains only high-quality, dynamically updated candidates.
Furthermore, we ensure that exactly $K$ nodes are chosen by enforcing the constraint $\sum_i(\mathbf{q}_i) = K$.
Beyond minimizing distribution shifts in GNN representations, our approach also reduces the KL distance between the homophily ratio distributions of the selection vector and the target.
\begin{align} \label{eq:optimization_improved}
\mathbf{q} = & \arg\min_\mathbf{q} \quad  \text{CMD}(\mathbf{Z}^{G}, \mathbf{q} \star \mathbf{Z}^{C}) + \lambda_{\text{S}} KL(\mathcal{B}^q, \mathcal{B}^{\text{target}})  \notag \\
& \text{s.t.} \quad \left\| \mathbf{q} \right\|_1 = K, \mathbf{q}_i \in [0, 1] \\
& |\mathcal{B}_i^q| = \sum_{v \in \mathcal{C}, j = \text{ind}(v, \mathcal{C}), \hat{h}(v) \in \left[\frac{i-1}{N}, \frac{i}{N}\right)} \mathbf{q}_j \notag
\end{align}
where $\mathbf{q}$ is the sole learnable parameter. CMD and KL metrics measure distribution distances, with $\mathcal{B}^q$ and $\mathcal{B}^{\text{target}}$ representing the homophily ratio distributions of the selection vector and the target, respectively. $\lambda_{\text{S}}$ controls homophily ratio consistency, $\mathbf{Z}^{G}$ is the global representations, $\mathbf{q} \star \mathbf{Z}^{C}$ computes weighted local representations of candidate pseudo-nodes, and $\text{ind}(v, \mathcal{C})$ indexes node $v$ in $\mathcal{C}$. After optimizing $\mathbf{q}$, the top $K$ nodes are selected as pseudo-nodes: 
\begin{equation}
    \mathcal{V}_{*,s}^P = \text{top-K}(\mathbf{q}, K, \mathcal{C})
    \label{eq:topK_pseudo_nodes}
\end{equation}
$\text{top-K}$ selects the $K$ highest-ranked nodes from $\mathcal{C}$ based on $\mathbf{q}$. This ensures consistency in GNN representation and homophily ratios.

\noindent\textbf{Loss Function of Pseudo-node Selection}. 
Our pseudo-node selection relies on a selection vector $\mathbf{q}$, which integrates the homophily ratio with the distribution consistency of GNN representations. The loss function for optimizing the selection vector is defined as: 
\begin{align} 
    \mathcal{L}_q = \text{CMD}(\mathbf{Z}^{G}, \mathbf{q} \star \mathbf{Z}^{C}) + \lambda_{\text{S}} KL(\mathcal{B}^q, \mathcal{B}^{\text{target}}) +  \max(0, \|\mathbf{q}\|_1 - K) 
    \label{eq:loss_q}
\end{align} 

\subsection{Pseudo-label Assignment with Multi-hop Neighbors}
After selecting pseudo-nodes, our framework assigns pseudo-labels. Traditional methods based on the highest prediction probability may be ineffective on heterophilic graphs, where GNNs relying on local neighbors often mislabel heterophilic nodes, increasing noise and exacerbating training bias. To address this, drawing from \cite{abu2019mixhop, zheng2022graph}, which show that higher-order neighbors more reliably identify similar nodes, we use these neighbors for heterophilic pseudo-nodes to improve labeling accuracy. For homophilic nodes, we continue using local one-hop neighbors.
The use of multi-hop neighbors is solely for assigning more accurate pseudo-labels; they are not employed in training the backbone GNN so as to preserve the original graph topology as including high-order neighbors can lose vital structural information, potentially affecting the estimation of homo ratios and calculation of distribution shifts in GNN representations.
To be specific, we define the \(k\)-hop adjacency matrix as \( \mathbf{A}^k\), the \(k\)-th power of \(\mathbf{A}\), with \(\mathbf{H} = f_\theta(\mathbf{X}, \mathbf{A}^k)\) for \(k \geq 2\)
representing the GNN's output in \(k\)-hop neighbors. The output used for pseudo-labeling is chosen based on the node homophily ratio: 
\begin{equation}
    \mathbf{Z}_i' =  \begin{cases}
        \mathbf{H}_i & \text{if } \hat{h}(v_i) < \delta_h, \\
        \mathbf{Z}^1_i & \text{if } \hat{h}(v_i) \geq \delta_h.
    \end{cases}
    \label{eq:multi_hop_Z}
\end{equation}
where \(\delta_h\) is the threshold for identifying heterophilic nodes. For the set of pseudo-nodes \(\mathcal{V}^P\), pseudo-labels are assigned based on the highest probability from the adaptive outputs:
\begin{equation}
    \mathbf{y}_i^{\text{pl}} = \arg \max_j \mathbf{Z}_{ij}', \quad v_i \in \mathcal{V}^P
    \label{eq:pseudo_labeling}
\end{equation}
This step is outlined in lines 9 to 10 of Algo. \ref{algo}. 

\subsection{Dual-Head GNN for Fully Utilizing Pseudo-nodes}
In graph self-training, training bias stems from the use of self-generated pseudo-labels. To mitigate this, we have introduced a careful selection of distribution-consistent pseudo-labeled nodes from a high-quality candidate set $\mathcal{C}$. However, this results in the discarding of many high-quality nodes that, while distribution-inconsistent, could be informative.
Our objective is dual: to reduce training bias and to enhance overall self-training performance. The latter goal is compromised by discarding potentially valuable pseudo-nodes. To this end, drawing inspiration from the image field where feature extractors and classifier heads in a neural network classifier are conceptually separated \cite{chen2022debiased}, we also apply a separation in our GNN model, consisting of a feature extractor $\psi$ and a classifier head $h_{\text{head}}$: 
\begin{equation}
f_\theta = h_{\text{head}} \circ \psi.
\end{equation}
To maximize the utility of all potential nodes, we employ a dual-head GNN training method. The main classifier head $h_{\text{main}}$ is trained on clean and carefully selected pseudo-nodes $\mathcal{V}^L \cup \mathcal{V}^P$. Meanwhile, the pseudo classifier head $h_{\text{pseudo}}$ enhances representation learning of the feature extractor by utilizing the remaining pseudo-nodes $\mathcal{C} - \mathcal{V}^P$, without affecting the performance of the main classifier head.
The training objective function is:
\begin{equation}
\min_{\psi, h, h_{\text{pseudo}}} \mathcal{L}_{\mathcal{V}^L \cup \mathcal{V}^P}(\psi, h_{\text{main}}) + \lambda_{\text{D}} \mathcal{L}_{\mathcal{C} - \mathcal{V}^P}(\psi, h_{\text{pseudo}})
\label{eq:dual_head_objective_function}
\end{equation}
where $h_{\text{pseudo}}$ independently processes distribution-inconsistent nodes. Though $h_{\text{main}}$ and $h_{\text{pseudo}}$ are fed with features from the same feature extractor, their parameters are independent. $\mathcal{L}$ is the cross entropy loss function. $\lambda_{\text{D}}$ is a control factor of the pseudo-head loss term. Importantly, the pseudo classifier head, only responsible for training the feature extractor, is not used during inference. This approach effectively balances reducing bias and enhancing model performance, ensuring the full utilization of all high-quality pseudo-nodes. This step is outlined in lines 11 to 13 of Algo. \ref{algo}.

\subsection{Workflow of HC-GST Framework}

After outlining the HC-GST framework's components—including distribution-consistent selection, multi-hop neighbors, and dual-head GNN—we detail our workflow in Fig. \ref{fig:framework} and Algo. \ref{algo}. Initially, we train a GNN on clean-labeled nodes to produce the output matrix \(\mathbf{Z}\). The pseudo-label set, \(\mathcal{V}_{*,s}^P\) (with \(s=1\) as the initial stage), begins empty. Nodes in the unlabeled set exceeding a confidence threshold \(\delta_c\) form a candidate set \(\mathcal{C}\), from which distribution-consistent pseudo-nodes, \(\mathcal{V}_{*,s}^P\), are selected. Next, a multi-hop neighbor method assigns pseudo-labels to these nodes. Finally, we replace the last GNN with a dual-head GNN trained on both the expanded labeled nodes and other high-quality nodes. We continuously update the pseudo-labeled node set to \(\mathcal{V}_{*, \leq s}^P\) with fresh labels and the output matrix with the new one, repeating these steps until we reach a predetermined number of stages \(S\) or achieve performance convergence. The GNN with the best validation accuracy is selected as the final model, with its output as the final result.

%% file: experiment.tex
\section{Experiments}
In this section, we conduct experiments on both homophilic and heterophilic graphs under various label rates to evaluate the effectiveness of the proposed framework.

\begin{table}[!t]
\centering
\caption{Statistics of datasets. $c$, $d$, and $h(\mathcal{G})$ are the number of classes, features, and graph homophily ratios, respectively. 
}
\vskip -1em 
\label{tab:dataset_stats}
\small
\begin{tabular}{lrrrrrr}
\hline
\textbf{Datasets} & $|\mathcal{V}|$    & $|\mathcal{E}|$    & $c$ & $d$    & $h(\mathcal{G})$ & Label Rate \\ \hline
Chameleon         & 2,277    & 31,421   & 5   & 2.325  & 0.23 & 1\%, 2\%, 5\% \\
Squirrel          & 5,201    & 198,493  & 5   & 2.089  & 0.22 & 1\%, 2\%, 5\% \\
Texas             & 183      & 295      & 5   & 1,703  & 0.11 & 5\%, 10\%, 30\% \\
arXiv-year        & 169,343  & 1,166,243 & 5  & 128    & 0.27 & 0.05\%, 0.1\% \\
Cora              & 3,327    & 4,676    & 7   & 3,703  & 0.74 & 1\% \\
Citeseer          & 2,708    & 5,278    & 3   & 1,433  & 0.81 & 1\% \\
PubMed            & 19,717   & 44,327   & 6   & 500    & 0.80 & 0.1\% \\ \hline
\end{tabular}
\vskip -1em
\end{table}

\subsection{Experimental Setup}




\subsubsection{Datasets} 
We use four heterophilic datasets: Chameleon, Squirrel, Texas, arXiv-year \cite{Rozemberczki2021, lim2021large}, and three homophilic datasets: Cora, Citeseer, PubMed \cite{bojchevski2018deep, sen2008collective, namata2012query}. Label rates $|\mathcal{V}^L|/|\mathcal{V}|$ vary from few-shot to standard; for example, $1\%$ in Chameleon equals $7$ labels per class. Texas ranges from $5\%$ ($2$ labels) to $30\%$ ($10$ labels) due to its small size. We set a $1\%$ label rate for homophilic graphs (PubMed 0.1\%). $0.5\%$ of nodes are randomly chosen as validation sets across all datasets, with the remainder as test nodes (or unlabeled nodes). Results are averaged over $10$ runs across all metrics and cases.


\subsubsection{Baselines} 
We compare HC-GST against four baselines. These include traditional self-training methods: (1) Naive \textbf{ST} \cite{li2018deeper}, selecting pseudo-nodes beyond a confidence threshold, and assigning the most confident label. (2) Cluster-aligned \textbf{M3S} \cite{sun2020multi}, choosing high-confidence pseudo-nodes and checking labeling precision via deep clustering. The other methods tackle distribution shifts: (3) \textbf{DRGST} \cite{liu2022confidence} selects high-confidence nodes, assigns the most confident label, and adjusts their weight by estimated information gain to mitigate distribution shifts. (4) \textbf{DCGST} \cite{wang2024DCGST} selects a proportion of high-confidence nodes to minimize distribution shifts using the CMD metric. All use a fixed backbone architecture.


\subsubsection{Experimental Details}
We use BMGCN~\cite{he2022block} as backbone GNN across all experiments and baselines. Key hyperparameters: $\lambda_{\text{S}}=2.0$ for homophily consistency, $\lambda_{\text{D}}=0.09$ for pseudo head modules, $\delta_c=0.65$ for confidence threshold, and $K$ matches initial labeled nodes. We utilize $k=2$ for two-hop neighbors, the heterophilic node threshold $\delta_h=0.4$, bins $N=10$, and stages $S=10$. Training uses a $0.001$ learning rate with Adam optimizer \cite{kingma2014adam}. We optimize hyperparameters for all methods using the validation set.

\subsubsection{Evaluation Metrics} \label{sec:metrics} 

The traditional accuracy metric provides a broad view of model performance but fails to measure nuances in training bias across various homophily bins during self-training. To capture these nuances, we introduce three metrics designed for evaluating training bias:
\begin{itemize}
    \item \textbf{Average Negative Performance Variation (\({NPV}\))}: This metric quantifies the average decrease in accuracy, capturing the detrimental impacts experienced during the self-training process as \({NPV} = \frac{1}{|\mathcal{N}_{-}|} \sum_{i\in \mathcal{N}_{-}} \text{ACC}_{\text{st},i} - \text{ACC}_{\text{backbone},i}\), 
    where \(\mathcal{N}_{-}\) is the set of bins with performance degradation, and \(\text{ACC}_{\text{st},i}\) and \(\text{ACC}_{\text{backbone},i}\) are the accuracies after self-training and from the original backbone model in the \(i\)-th bin respectively.

    \item \textbf{Average Positive Performance Variation (\({PPV}\))}: This metric averages the positive increments in accuracy, highlighting the beneficial effects of self-training as \({PPV} = \frac{1}{|\mathcal{N}_{+}|} \sum_{i \in \mathcal{N}_{+}} \text{ACC}_{\text{st},i} - \text{ACC}_{\text{backbone},i}\), 
    where \(\mathcal{N}_{+}\) is the set of bins with performance improvements.

    \item \textbf{Total Performance Variation (\({TPV}\))}: This metric averages the overall variations in accuracy across all bins, providing an overall measure of self-training impact as \({TPV} = \frac{1}{N} \sum_{i=1}^{N} \text{ACC}_{\text{st},i} - \text{ACC}_{\text{backbone},i}\).
\end{itemize}
These metrics offer a detailed view of performance dynamics across homophily bins, enhancing our understanding of how self-training methods affect model behavior. The ACC and \( TPV \) are comprehensive metrics that provide a broad overview of performance.


\begin{table*}[!htbp]
\centering
\caption{Comparison results (\%) on biased training samples across heterophilic graph datasets. (bold: best)}
\vskip -1em 
\label{tab:comparison_biased_hetero_dataset}
\small
\begin{tabular}{c|c|c|c|c|c|c|c|c|c|c|c|c|c}
\hline
 \multirow{2}{*}{Method}&  & \multicolumn{3}{c|}{Chameleon} & \multicolumn{3}{c|}{Squirrel} & \multicolumn{3}{c|}{Texas} & \multicolumn{2}{c|}{arXiv-year} & \multirow{2}{*}{Average} \\
\cline{2-13}
 & Label Rate & 5 & 10 & 20 & 10 & 20 & 50 & 2 & 4 & 10 & 17 & 34 & \\
\hline
BMGCN  & ACC & 31.67 & 38.23 & 45.65 & 22.13 & 24.43 & 28.32 & 30.79 & 35.43 & 48.37 & 30.56 & 31.99 & 33.51 \\
\hline
ST     & ACC & 31.73 & 38.74 & 45.08 & 23.00 & 25.90 & 27.74 & 31.95 & 42.42 & 50.72 & 31.32 & 32.10 & 34.01 \\
       & ${TPV}$ & -0.88 & -3.71 & -0.73 & \textbf{4.14} & 0.70 & -0.39 & -1.72 & 2.04 & -5.52 & -5.18 & -2.53 & -1.02 \\
       & ${NPV}$ & -4.83 & -6.45 & -4.15 & -2.14 & -3.45 & \textbf{-2.89} & -25.99 & -34.87 & -33.80 & -8.99 & -7.20 & -12.22 \\
       & ${PPV}$ & 1.97 & 0.38 & 1.86 & 8.16 & 4.18 & 3.08 & 7.99 & 11.81 & 1.53 & 3.70 & 2.14 & 4.22 \\
\hline
M3S    & ACC & 33.87 & 38.09 & 45.59 & 22.15 & 23.63 & 27.56 & 35.40 & 38.41 & 50.55 & 30.92 & 33.79 & 34.21 \\
       & ${TPV}$ & 0.92 & 0.31 & -1.24 & -0.07 & -3.59 & -0.82 & 0.41 & \textbf{4.70} & \textbf{0.63} & 2.04 & 2.99 & 0.57 \\
       & ${NPV}$ & -4.19 & -3.71 & -4.76 & -0.51 & -5.44 & -3.21 & -20.86 & -20.26 & \textbf{-14.38} & -1.35 & -1.24 & -8.02 \\
       & ${PPV}$ & 5.08 & 4.62 & 2.71 & 0.21 & 2.20 & 2.82 & 3.28 & 7.00 & 2.45 & 3.50 & 4.05 & 3.38 \\
\hline
DRGST  & ACC & 34.18 & 37.36 & 43.74 & 21.78 & 22.68 & 25.34 & 29.20 & 25.78 & 28.38 & 30.78 & 32.69 & 29.91 \\
       & ${TPV}$ & -5.44 & -0.41 & -1.15 & 2.85 & -2.65 & -0.11 & -1.70 & 0.52 & -11.79 & -0.81 & 1.84 & -1.57 \\
       & ${NPV}$ & -15.51 & -4.74 & -6.05 & -5.16 & -7.22 & -6.49 & -24.67 & -40.99 & -42.18 & -1.89 & -0.98 & -13.96 \\
       & ${PPV}$ & 6.71 & 1.43 & 2.98 & 9.88 & 7.17 & 11.87 & 9.71 & 22.24 & 19.32 & 1.69 & 6.09 & 8.72 \\
\hline
DCGST  & ACC & 34.79 & 40.63 & 44.10 & 22.96 & 22.95 & 24.37 & 40.75 & 27.37 & 32.09 & 30.59 & 32.05 & 32.81 \\
       & ${TPV}$  & 1.46 & 0.23 & -0.63 & -0.58 & -0.29 & -1.36 & \textbf{4.05} & -3.85 & -5.84 & -0.93 & 0.16 & -0.63 \\
       & ${NPV}$ & -4.84 & -3.78 & -3.81 & -5.19 & -3.98 & -4.95 & -27.06 & -37.27 & -31.35 & -2.87 & -0.51 & -11.34 \\
       & ${PPV}$ & 8.24 & 3.56 & 0.92 & 2.87 & 3.59 & 4.01 & 12.28 & 7.80 & 12.03 & 0.99 & 0.84 & 4.94 \\
\hline
Ours   & ACC & \textbf{36.36} & \textbf{42.74} & \textbf{46.11} & \textbf{25.15} & \textbf{26.68} & \textbf{29.70} & \textbf{42.57} & \textbf{45.20} & \textbf{52.39} & \textbf{33.36} & \textbf{35.67} & \textbf{37.95} \\
       & ${TPV}$ & \textbf{4.22} & \textbf{2.24} & \textbf{1.53} & 1.79 & \textbf{5.93} & \textbf{1.57} & 0.54 & 4.25 & -2.75 & \textbf{3.52} & \textbf{10.20} & \textbf{2.68} \\
       & ${NPV}$ & \textbf{-2.46} & \textbf{-0.24} & \textbf{-0.71} & \textbf{-0.31} & \textbf{-1.57} & \textbf{-2.89} & \textbf{-18.72} & \textbf{-17.03} & -24.24 & \textbf{-1.12} & \textbf{-0.41} & \textbf{-6.10} \\
       & ${PPV}$ & \textbf{6.48} & \textbf{2.86} & \textbf{0.96} & \textbf{2.69} & \textbf{8.02} & \textbf{2.34} & \textbf{8.80} & \textbf{10.49} & \textbf{2.61} & \textbf{8.98} & \textbf{12.85} & \textbf{6.11} \\
\hline
\end{tabular}
\end{table*}

\subsection{Results and Analysis} 
To explore HC-GST’s capability with a biased training set in homophily ratio distribution, we randomly generate local homophily ratio distributions for the training set that deviates from the global distribution. Nodes are then selected one by one according to the local distribution.
Tables \ref{tab:comparison_biased_hetero_dataset} and \ref{tab:comparison_biased_homo_dataset} compare our method with baseline approaches on datasets with biased training sets in both homophily and heterophily settings. We observe: (1) Across four heterophily graphs, our method excels in comprehensive metrics, ACC and $TPV$, improving by $3.74\%$ and $2.11\%$ respectively over the runner-up on average. This highlights the effectiveness of our framework in self-training on heterophilic graphs. (2) For training bias metrics, $NPV$ and $PPV$ (where a smaller gap is preferable), our method consistently outperforms others, showing that our distribution consistency selection effectively reduces the training bias. Conversely, ST and M3S ignore distribution shifts, leading to significant training bias. DRGST and DCGST, despite considering consistency, fail on heterophilic graphs due to biased training nodes that significantly affect GNN representations. (3) We also note similar patterns across three homophilic graphs. This indicates that homophily ratio distribution shifts are a widespread issue in graph learning, irrespective of the graph type. Our best performance proves the generalization of our method across heterophily and homophily settings. (4) Examining accuracy changes under various label rates between our method and the backbone, we observe more substantial improvements at lower label rates. This indicates that our self-training approach is particularly effective in few-shot scenarios.

Table \ref{tab:comparison_random_dataset} details results on six graphs at a $1\%$ label rate without obvious homophily ratio distribution shift between the clean training set and the global graph. This is achieved through normal random sampling of training nodes, which tends to produce a homophily ratio distribution similar to the global one. Negligible shifts in the training set suggest that distribution shifts primarily arise from pseudo-labeling. Compared to DRGST and DCGST, which focus on reducing distribution shifts through GNN representations, our method is more effective across all metrics, demonstrating the effectiveness of our homophily ratio consistency module. In contrast, ST and M3S exhibit the most significant training bias, underscoring the importance of addressing distribution shifts during pseudo-labeling.

\begin{table}[!htbp]
\centering
\caption{Results (\%) on biased training samples across homophilic graph datasets at $1\%$ label rate (PubMed $0.1\%$). }
\vskip -1em 
\label{tab:comparison_biased_homo_dataset}
\small
\begin{tabular}{c|c|c|c|c|c}
\toprule
\multirow{1}{*}{Method} & \multicolumn{1}{c|}{Metric} & \multicolumn{1}{c|}{Cora} & \multicolumn{1}{c|}{Citeseer} & \multicolumn{1}{c|}{PubMed} & \multicolumn{1}{c}{Average} \\ 
\cline{1-6}
BMGCN & ACC & 52.24 & 55.63 & 68.83 & 58.90 \\
\hline 
ST & ACC & 55.72 & 61.16 & 70.87 & 62.58 \\
 & ${TPV}$ & 2.19 & 1.13 & 2.40 & 1.91 \\
 & ${NPV}$ & -4.59 & -1.83 & -3.62 & -3.35 \\
 & ${PPV}$ & 4.39 & 2.18 & 4.02 & 3.53 \\
\hline
M3S & ACC & 58.26 & 60.68 & 70.76 & 63.23 \\
 & ${TPV}$ & 4.65 & 0.94 & 2.10 & 2.56 \\
 & ${NPV}$ & -1.40 & -2.60 & -4.73 & -2.91 \\
 & ${PPV}$ & 5.46 & 1.90 & 4.36 & 3.91 \\
\hline
DRGST & ACC & 57.22 & 58.21 & 71.52 & 62.32 \\
 & ${TPV}$ & 1.72 & 1.04 & 1.08 & 1.28 \\
 & ${NPV}$ & -4.45 & \textbf{-0.69} & -6.84 & -3.99 \\
 & ${PPV}$ & 3.55 & 1.37 & 4.49 & 3.14 \\
\hline
DCGST & ACC & 58.19 & 60.77 & 70.19 & 63.05 \\
 & ${TPV}$ & 1.35 & 2.91 & 1.00 & 1.75 \\
 & ${NPV}$ & -5.26 & -3.65 & \textbf{-2.11} & -3.67 \\
 & ${PPV}$ & 5.04 & 3.53 & 3.08 & 3.88 \\
\hline
Ours & {ACC} & \textbf{58.94} & \textbf{64.85} & \textbf{72.68} & \textbf{65.49} \\
 & \textbf{${TPV}$} & \textbf{5.50} & \textbf{5.62} & \textbf{3.49} & \textbf{4.87} \\
 & \textbf{${NPV}$} & \textbf{-0.96} & -1.39 & -3.67 & \textbf{-2.01} \\
 & \textbf{${PPV}$} & \textbf{6.15} & \textbf{6.11} & \textbf{5.91} & \textbf{6.06} \\
\bottomrule
\end{tabular}
\end{table}

\begin{table}[!htbp]
\centering
\caption{Results (\%) on random training samples across graph datasets at $1\%$ label rate (Texas $5\%$, PubMed $0.1\%$). }
\vskip -1em 
\label{tab:comparison_random_dataset}
\small
\newcolumntype{P}[1]{>{\centering\arraybackslash}p{#1}}
\resizebox{1.00\linewidth}{!}{%
\begin{tabular}{c|P{0.65cm}|P{0.65cm}|P{0.65cm}|P{0.7cm}|P{0.65cm}|P{0.65cm}|P{0.65cm}|P{0.65cm}}
\toprule
{Method} & &  {\footnotesize \shortstack{Cham.}} & {\footnotesize Squi.} & Texas & Cora & {\footnotesize Cite.} & {\footnotesize Pubm.} & Avg. \\ 
\hline
\multirow{1}{*}{BMGCN} & ACC & 34.74 & 24.21 & 40.51 & 58.22 & 47.96 & 70.55 & 46.03 \\
\hline
\multirow{5}{*}{ST} & ACC & 34.90 & 23.74 & 39.48 & 61.36 & 49.47 & 71.74 & 46.78 \\
 & ${TPV}$ & 0.72 & 1.48 & 1.89 & 1.43 & 1.99 & 0.21 & 1.12 \\
 & ${NPV}$ & -5.76 & -5.80 & -25.94 & -1.71 & -1.87 & -2.08 & -7.19 \\
 & ${PPV}$ & \textbf{6.00} & \textbf{4.28} & 11.75 & 2.57 & 3.60 & 1.26 & 4.91 \\
\midrule
\multirow{5}{*}{M3S} & ACC & 35.58 & 23.23 & 40.51 & 59.62 & 49.43 & 71.68 & 46.68 \\
 & ${TPV}$ & 1.27 & -0.07 & -3.50 & 1.16 & 1.36 & 0.18 & 0.07 \\
 & ${NPV}$ & -1.42 & -1.93 & -32.29 & -0.51 & \textbf{-1.42} & -2.04 & -6.60 \\
 & ${PPV}$ & 2.89 & 1.79 & \textbf{32.42} & 1.57 & 2.89 & 1.09 & 7.11 \\
\midrule
\multirow{5}{*}{DRGST} & ACC & 35.75 & 24.79 & 42.01 & 62.82 & 49.53 & 71.28 & 47.70 \\
 & ${TPV}$ & 0.36 & -1.85 & -3.10 & 3.13 & 1.39 & 0.12 & 0.01 \\
 & ${NPV}$ & -4.13 & -3.93 & -18.25 & -9.34 & \textbf{-1.42} & -2.69 & -6.63 \\
 & ${PPV}$ & 2.75 & 4.05 & 6.84 & \textbf{5.05} & 2.06 & 1.79 & 3.76 \\
\midrule
\multirow{5}{*}{DCGST} & ACC & 35.25 & 25.24 &\textbf{ 47.92} & 61.32 & 57.61 & 73.18 & 50.09 \\
 & ${TPV}$ & 0.44 & -1.64 & 1.77 & 1.55 & 2.11 & 1.03 & 0.88 \\
 & ${NPV}$ & -4.81 & -4.87 & -12.36 & -3.14 & -4.52 & \textbf{-1.44} & -5.19 \\
 & ${PPV}$ & 5.08 & 3.09 & 6.10 & 2.87 & 3.42 & 1.53 & 3.68 \\
\midrule
\multirow{5}{*}{Ours} & {ACC} & \textbf{36.54} & \textbf{25.42} & 44.41 & \textbf{66.97} & \textbf{61.33} & \textbf{74.04} & \textbf{51.62} \\
 & \textbf{${TPV}$} & \textbf{2.98} & \textbf{0.56} & \textbf{3.96} & \textbf{4.66} & \textbf{6.91} & \textbf{1.62} & \textbf{3.45} \\
 & \textbf{${NPV}$} & \textbf{-0.63} & \textbf{-1.42} & \textbf{-9.16} & \textbf{-0.17} & -3.58 & -1.75 & \textbf{-2.79} \\
 & \textbf{${PPV}$} & 3.50 & 2.01 & 26.28 & 4.82 & \textbf{10.29} & \textbf{2.81} & \textbf{8.29} \\
\bottomrule
\end{tabular}
}
\end{table}


\subsection{Ablation Study}


To assess the impact of each component within the HC-GST framework, we compare it against various variants: (1) HCGST-w/o-MultiHop, removing the multi-hop neighbors module; (2) HCGST-w/o-DualHead, without the dual-head mechanism; (3) HCGST-w/o-Selection, which excludes the distribution consistency selection. For clarity, we remove the hyphen from our framework name. We conduct this ablation study on Chameleon with a biased training set at a $1\%$ label rate. As shown in Table \ref{tab:ablation_study}, the findings are: (1) Removing the distribution consistency module leads to a significant accuracy decline of $4.34\%$, underscoring the importance of distribution consistency in selecting high-quality pseudo-labels. Besides, the gap between $NPV$ and $PPV$ is larger than in the original, illustrating its contribution to mitigating training bias; (2) The absence of the multi-hop neighbors results in a $1.94\%$ drop in accuracy, validating its importance for accurately assigning pseudo-labels; (3) Without the dual-head mechanism, accuracy decreases by $0.61\%$, underscoring its effectiveness in fully utilizing pseudo-nodes.


\begin{table}[!ht]
\centering
\caption{Ablation study results (\%) for different components.}
\vskip -1em 
\label{tab:ablation_study}
\begin{tabular}{l|c|c|c|c}
\toprule
\textbf{Model Variant} & ACC  & ${TPV}$ & ${NPV}$ & ${PPV}$ \\ \hline
HCGST                   & 36.36 & 4.22 & -2.46 & 6.48  \\ 
HCGST-w/o-DualHead         & 35.75 & 1.12 & -2.19 & 3.57  \\ 
HCGST-w/o-MultiHop             & 34.42 & 0.86 & -1.70 & 4.72  \\ 
HCGST-w/o-Selection   & 32.02 & -0.43 & -4.57 & 6.19  \\ 
Backbone               & 31.67 &   - & - & -     \\ 
\bottomrule
\end{tabular}
\end{table}

\subsection{Analysis of Hyper-parameter Impact}

\begin{figure}[!t]
    \centering
    \includegraphics[width=1.\linewidth]{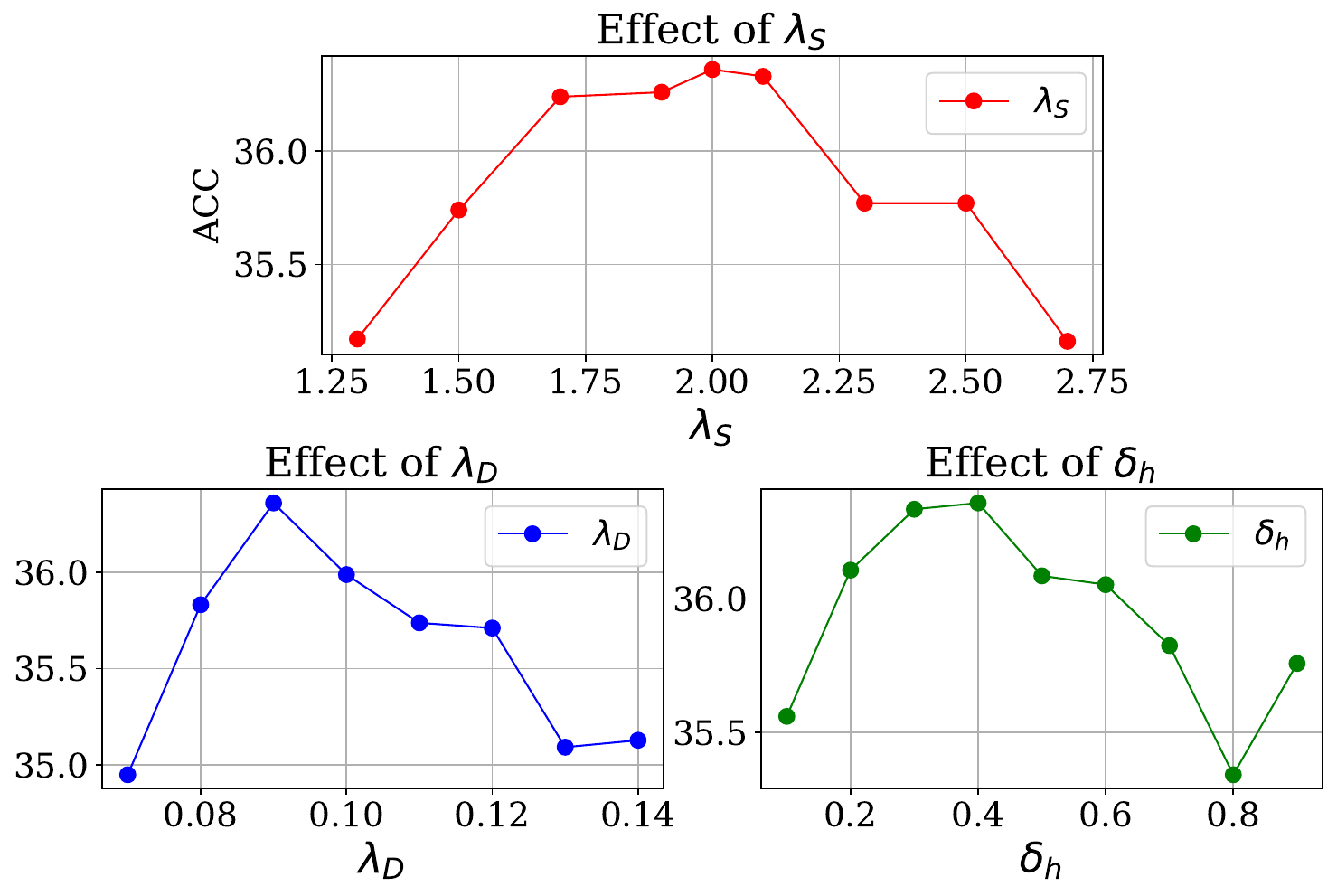}
    \vskip -1em
    \caption{Sensitivity analysis of hyper-parameters w.r.t. $\lambda_{\text{S}}$, $\lambda_{\text{D}}$, and $\delta_{h}$ on Chameleon with $1\%$ label rate. }
    \label{fig:hyperparameter}
    \vskip -1em
\end{figure}

We assess the impacts of key hyperparameters $\lambda_{\text{S}}$, $\lambda_{\text{D}}$, and $\delta_{h}$ within the HC-GST framework, which regulate the distribution consistency selection module, the dual-head module, and the multi-hop neighbors module, respectively. To explore $\lambda_{\text{S}}$, we fix $\lambda_{\text{D}} = 0.09$ and $\delta_{h} = 0.4$, and vary $\lambda_{\text{S}}$ between $\{1.3, 1.5, \dots, 2.7\}$. Similarly, with $\lambda_{\text{S}} =2.0$ and $\delta_{h} = 0.4$, we vary $\lambda_{\text{D}}$ from $\{0.07, 0.08, \dots, 0.14\}$, and with $\lambda_{\text{S}} =2.0$ and $\lambda_{\text{D}} = 0.09$, we adjust $\delta_{h}$ between $\{0.1, 0.2, \dots, 0.9\}$. As shown in Fig. \ref{fig:hyperparameter}: (1) Our model is robust to minor deviations from optimal hyperparameter values but underperforms with extreme values; (2) The optimal $\lambda_{\text{S}}$ is around $2.0$, with higher values leading to deviations from GNN representation-based distribution, adversely affecting performance; (3) The best $\lambda_{\text{D}}$ is near $0.09$, suggesting moderate utilization of pseudo-nodes beyond carefully-selected ones aids in training feature extractor, and higher values may introduce noise; (4) An ideal $\delta_{h}$ of about $0.4$ indicates a trade-off between selecting one-hop and multi-hop neighbors, with excessive reliance on multi-hop neighbors decreases the accuracy of assigning pseudo-labels for homophilic nodes and lower performance.

\subsection{Benefits of Reducing Heterophily-aware Distribution Shift}
\label{sec:visual}

\begin{figure}[!htbp]
    \centering
    \includegraphics[width=1.\linewidth]{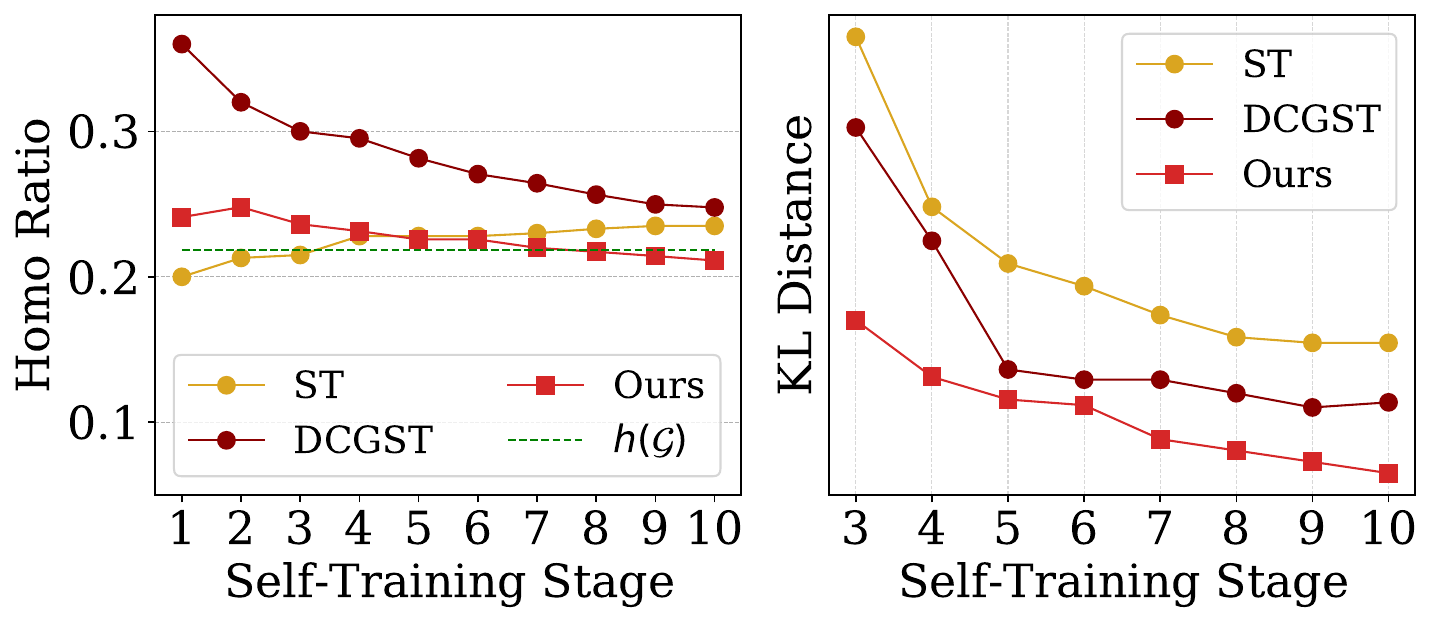}
    \vskip -1em
    \caption{Left: mean homophily ratio of pseudo-nodes during self-training stages on Squirrel. Right: KL divergence between local and global homophily ratio distributions.}
    \label{fig:homo_ratio_distance}
    \vskip -1em
\end{figure}

\begin{figure}[!htbp]
    \centering
    \includegraphics[width=0.92\linewidth]{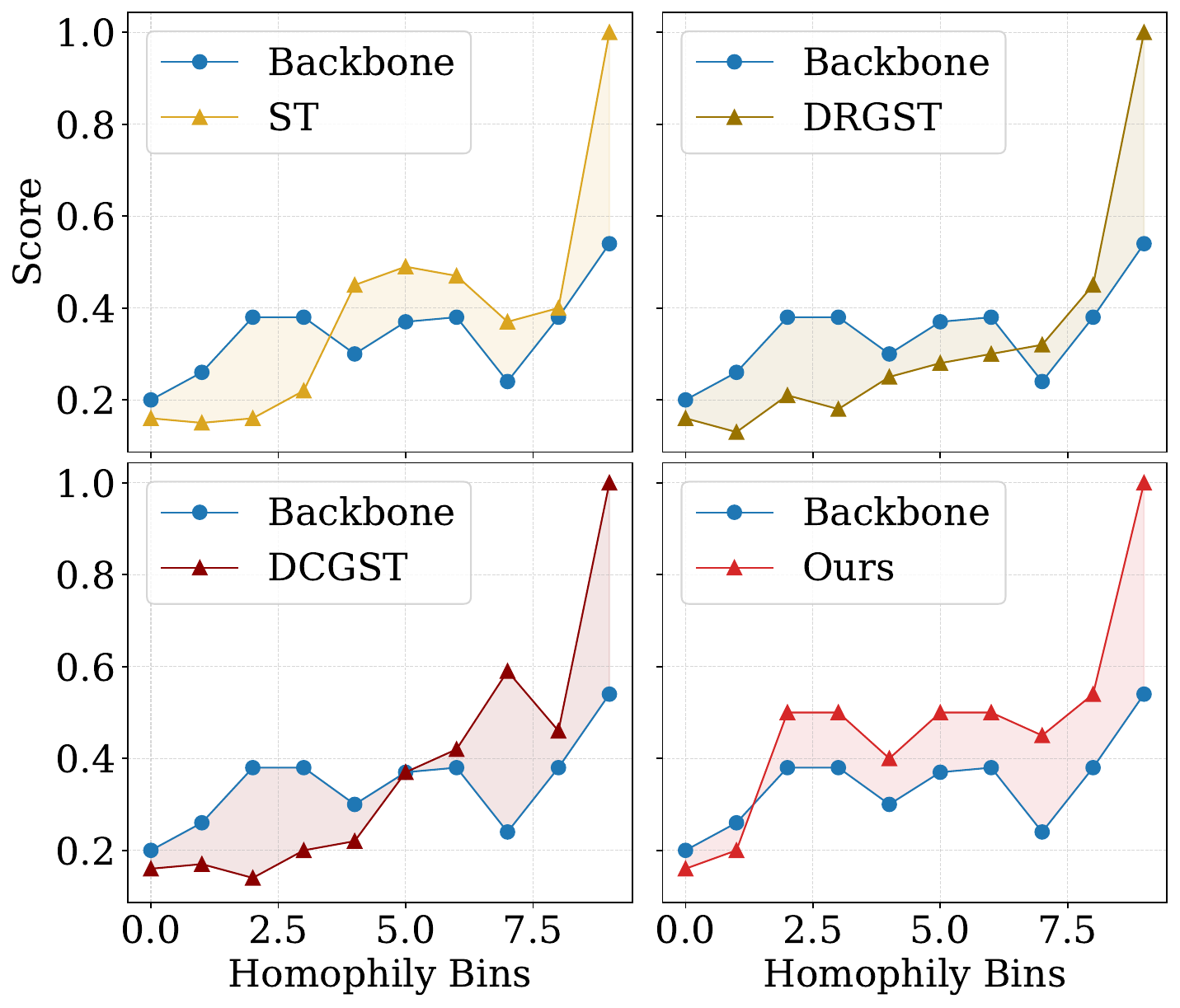}
    \vskip -1em
    \caption{Training bias across various homophily bins on Chameleon graph with $1\%$ label rate.  }
    \label{fig:training_bias}
    \vskip -1em
\end{figure}

To highlight the benefits of reducing shifts in homophily ratio distribution, we visualize changes in homophily ratios on Squirrel and training bias on Chameleon at a $1\%$ label rate. Fig. \ref{fig:homo_ratio_distance} displays variations in mean homophily ratios and the KL divergence between local and global distributions during self-training stages. Fig. \ref{fig:training_bias} compares accuracies across different homophily bins before and after the baselines and with our method. Our findings include: (1) Our method selects pseudo-nodes closer to the global distribution more effectively than baselines, as evidenced by closer mean homophily ratio values and reduced KL distance in Fig. \ref{fig:homo_ratio_distance};  (2) Consistent homophily between pseudo-nodes and the global distribution reduces training bias, as indicated by our smaller negative and larger positive areas than baselines across homophily bins in Fig. \ref{fig:training_bias}.


